# First Proof of Principle Experiment for Muon Production with Ultrashort High Intensity Laser

**Authors:** Feng Zhang[1]*, Li Deng[2,3]*, Yanjie Ge[4,9]*, Jiaxing Wen[1], Bo Cui[1], Ke Feng[4], Hao Wang[4], Chen Wu[5,6], Ziwen Pan[7], Hongjie Liu[1], Zhigang Deng[1], Zongxin Zhang[4], Liangwen Chen[3,2,8]†, Duo Yan[3,2,8], Lianqiang Shan[1], Zongqiang Yuan[1], Chao Tian[1], Jiayi Qian[4], Jiacheng Zhu[4], Yi Xu[4], Yuhong Yu[3,2,8], Xueheng Zhang[3,2,8], Lei Yang[3,2,8], Weimin Zhou[1]†, Yuqiu Gu[1]†, Wentao Wang[4]†, Yuxin Leng[4], Zhiyu Sun[3,2,8], Ruxin Li[4]

**Affiliations:**

[1] Nation Key Laboratory of Plasma Physics, Laser Fusion Research Center (LFRC), Academy of Engineering Physics (CAEP), Mianyang 621900, China

[2] Advanced Energy Science and Technology Guangdong Laboratory, Huizhou 516000, China

[3] Institute of Modern Physics, Chinese Academy of Sciences (CAS), Lanzhou730000, China

[4] State Key Laboratory of High Field Laser Physics and CAS Center for Excellence in Ultra-intense Laser Science, Shanghai Institute of Optics and Fine Mechanics (SIOM), Chinese Academy of Sciences (CAS), Shanghai 201800, China

[5] Institute of High Energy Physics, Chinese Academy of Sciences (CAS), Beijing 100049, China

[6] Spallation Neutron Source Science Center, Dongguan 523808, China

[7] State Key Laboratory of Particle Detection and Electronics, University of Science and Technology of China, Hefei 230026, China

[8] School of Nuclear Science and Technology, University of Chinese Academy of Sciences, Beijing 100049, China

[9] Center of Materials Science and Optoelectronics Engineering, University of Chinese Academy of Sciences, Beijing 100049, China.

[†]Correspondence author. Email:   zhouwm@caep.cn, yqgu@caep.cn, chenlw@impcas.ac.cn, wwt1980@siom.ac.cn.

*These authors contributed equally to this work.

**Abstract:**

  Muons, which play a crucial role in both fundamental and applied physics, have traditionally been generated through proton accelerators or from cosmic rays. With the advent of ultra-short high-intensity lasers capable of accelerating electrons to GeV levels, it has become possible to generate muons in laser laboratories. In this work, we show the first proof of principle experiment for novel muon production with an ultra-short, high-intensity laser device through GeV electron beam bombardment on a lead converter target. The muon physical signal is confirmed by measuring its lifetime which is the first clear demonstration of laser-produced muons. Geant4 simulations were employed to investigate the photo-production, electro-production, and Bethe-Heitler processes response for muon generation and their subsequent detection. The results show that the dominant contributions of muons are attributed to the photo-production/electro-production and a significant yield of muons up to 0.01 μ/e⁻ out of the converter target could be achieved. This laser muon source features compact, ultra-short pulse and high flux. Moreover, its implementation in a small laser laboratory is relatively straightforward, significantly reducing the barriers to entry for research in areas such as muonic X-ray elemental analysis, muon spin spectroscopy and so on.

**Text:**

Laser devices are among the most powerful machines created by humans, capable of generating extreme material states to those inside stars in short periods [1], achieving controlled nuclear fusion in laboratories [2] and producing a wide variety of radiation sources such as X-ray and neutron sources [3], high-intensity electron beams [4], high energy proton beams (up to 150 MeV) [5] and electron beams (up to several GeV) [6]. Those electron and proton beams could led to the development of various secondary radiation sources, including betatron radiation [7], positrons [8], bremsstrahlung sources [9], neutron sources [10], and free electron lasers [11]. As the energy scale of laser radiation sources increases, new types of radiation sources, such as muons, have become possible.

The muon, an elementary particle in the standard model of particle physics, has a rest mass of 106 MeV/$c^2$ and an average lifetime of 2.2 μs. Its decay produces an electron and two neutrinos. Historically, muons played a pivotal role in fundamental physics research. Recent advancements in the muon g-2 precision measurement have revealed significant findings that suggest the emergence of new physics [12]. Muons also play an important role in probing Charged Lepton Flavor Violation processes, the detection of which would be a clear signature of new physics beyond Standard Model [13].With muons being approximately 207 times more massive than electrons, muon colliders can be considerably more compact than electron colliders, making them a processive focus in high-energy physics [14, 15]. The strong penetrative abilities of muons allow their use in imaging technologies and other fields, as demonstrated in research on muon radiography [16, 17]. Muon Spin Rotation/Relaxation/Resonance (µSR) is an invaluable tool for investigating the static and dynamical magnetic properties of materials [18]. Additionally, muons can act as catalysts in cold fusion reactions [19]. Evidently, muon physics remains a central area of both fundamental and applied physics research.

Traditionally, muons are predominantly generated by accelerating protons to energies of several hundred MeV. These protons collide with carbon or other targets to produce pions or kaons, which then decay into muons. Currently, there are several operational muon sources globally [20-24]. Additionally, facilities such as SNS, CSNS, HIAF/CiADS, SHINE, RAON, FNAL have the capability to provide high intensity muons [25-30]. However, these sources are characterized by fixed locations, substantial investment costs, and high operating expenses. An alternative source of muons comes from cosmic rays [31], which are promising for applications like muon imaging. Despite their low flux (approximately 1/cm²/min) and high average energy, it imposes certain limitations on research.

The energy scale of ultra-short, ultra-intense lasers now exceeds the threshold required for muon generation, making it feasible to produce muons using GeV-level electrons. In 2009, A.I. Titov et al. [32] proposed using LWFA electron beams to generate $\mu^+\mu^-$ pairs via the Bethe-Heitler process. Further studies have optimized conditions for muon pair production [33]. Notably, muons can also be produced through photo-production or electro-production, which have significantly higher cross sections compared to the Bethe-Heitler process [34].

Muons generated this way exhibit short duration and high flux, naturally synchronizing with laser facilities, making them excellent for investigating extreme material states induced by lasers. However, empirical confirmation of muon production in laser laboratories remains elusive. The main challenge is the relatively low cross section for muon production, alongside the simultaneous generation of various radiations (e.g., gamma rays, electrons, neutrons et al.,) within the conversion target, which leads to detector saturation and obscures muons. In 2017, we proposed a method to diagnose muons generated from this scheme by measuring the muon lifetime [35]. The ultrashort pulse duration (~tens of *fs*) of the LWFA electrons allows us to assume that the muons are generated concurrently. By capturing the decay electron/positron in detectors, we can confirm the muon signal by analyzing their lifetimes. This technique has high signal discrimination capabilities and supports multiple accumulations, making it particularly suitable for scenarios where the individual yield in a single shot is not very high. Moreover, considering that the average decay time of muons is 2.2 μs, the instantaneous radiation produced during the shooting process significantly decreases after this period, simplifying the identification of muons.

In this paper, we present the first proof of principle experiment for muon production with an ultra-short, high-intensity laser device through LWFA electron beam bombardment on a lead (Pb) converter target. We detected muon and measured its lifetime which is the first clear demonstration of laser-produced muons. We also evaluated the detection efficiency of our measurement system using global Geant4 simulation [36], determining that the muon yield reaches $0.01\mu/e^-$ with an instantaneous signal strength of up to $10^7$ in a single shot.

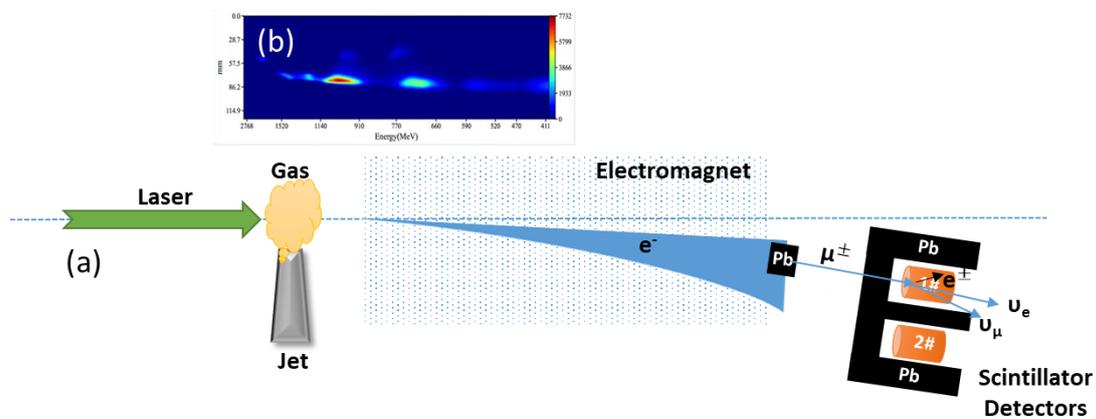

**Figure 1 Schematic of the experimental setup.** (a) A 500-660 TW, 30-fs laser pulse was focused onto a gas jet to accelerate electron beams. The spectra of the electron beams were measured using a 1-Tesla, 2-meter-long magnetic spectrometer. Subsequently, the electron beams entered a 12-cm-thick Pb converter target to generate muon. Muons were then collected by two scintillator detectors. (b) The typical energy spectrum after the overlay of all electron beams.

The experiment was conducted at the 1 Petawatt laser system of the Shanghai Ultra-Intensive Ultra-Short Laser Experiment Facility (SULF) [37]. As illustrated in Figure 1(a), the laser energy delivered to the target ranged from 15-20J, with a central wavelength of ~800 nm, operating at a repetition rate of 0.1 Hz to accelerate high-energy electron beams (see Methods). A total of 178 shots were performed in the experiment. The energy range of the electron beams, obtained from the electron spectrometer, was between 0.4 and 1.5 GeV with an average charge of ~200 pC. The energy spectra of all electron beams were superimposed, as depicted in Fig. 1(b), corresponding to a total charge of 40 nC and an electron yield of $2\times10^{11}$. Subsequently, the electron beams were directed towards the conversion target by an electromagnet (1.0 Tesla) for collision with the Pb converter to generate muons. A 12 cm thick Pb block with a cross-section of 10cm×10cm was used as the conversion target to optimize the retention of gamma photons produced by electron collisions and improve the muon conversion efficiency.

Two liquid scintillator detectors were positioned behind the conversion target to detect positrons/electrons decayed from muons during the experiment (see Methods). These detectors, oriented towards the direction of muons, had a cross-sectional diameter of Φ14cm and a thickness of 10 cm, with a 0.6 cm Al shell cladding. The detectors were shielded by Pb with a thickness of 9 cm in the muon-facing direction and 5 cm in lateral directions. As shown in Figure 1, Detector 1# covered an angular range of ±13°, while Detector 2# spanned from 22° to 41°. The liquid scintillator detectors were equipped with Microchannel Plate (MCP) photomultiplier tubes (MCP-PMT) and incorporated gating functionality. MCPs could experience saturations due to the instantaneous and intense ionizing radiation produced by electron beams hitting the conversion target. To prevent signal saturations, the MCP-PMTs were triggered 2 μs after the target shot within an acquisition time window to 20 μs.

The electron beam and its bremsstrahlung photons instantly generate muon pairs through the Bethe-Heitler process. Due to the short distance from the conversion target to the detector, it is reasonable to assume that muon generation coincides with the moment when the instantaneous radiation from the shot reaches the detector, which is defined as the starting time of muon generation, denoted as $t_{start}$. Additionally, muons are also generated from the decay of mesons which are produced by photo-production and electro-production [34]. Considering that the lifetime of pion is much shorter than that of muon, it is reasonable to assume muons are produced simultaneously. Therefore, muons from the three processes are integrally analyzed in this work. Those generated muons come to a stop within the lead shielding wall or the detectors themselves, decaying into electrons/pisitrons and neutrinos. While neutrinos can freely exit the detection area, electrons are detectable by the MCP-PMTs, indicating the muon decay time, denoted as $t_{stop}$. From the distribution of the time difference $\Delta t = t_{stop} - t_{start}$, the lifetime of the generated muons can be analyzed by comparing to the known distribution of muon decay times.

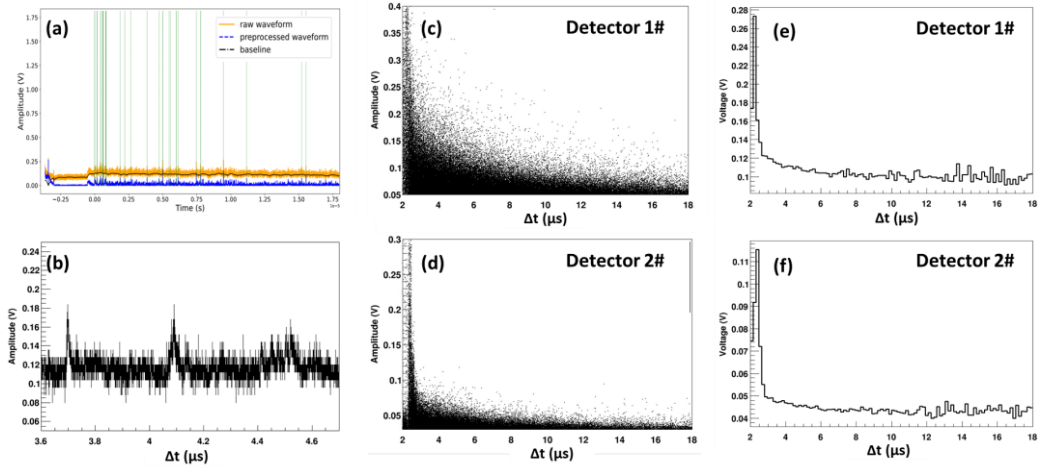

**Figure 2 The distribution of the time difference $\Delta t$ in the experiment.** (a) The typical data recorded by a 2.5GHz oscilloscope in a single shot, subtracting the baseline (black curve) from the raw data (yellow curve) yields the waveform identified as a muon decay event (blue curve). Each green vertical line in the figure indicates an instance of muon decay. (b) Waveform of a typical muon decay signal peak out of the MCP-PMT. (c) and (d) The two-dimensional scatter plot of amplitude versus $\Delta t$ for Detector 1# and 2#. (e) and (f) The average signal amplitude per unit time as a function of $\Delta t$.

The electrical pulse signals out of MCP-PMT were captured by a 2.5GHz oscilloscope, with typical data from a single shot shown in Figure 2(a). The sharp peak on the far left in Figure2(a) corresponds to the instantaneous radiation signal generated when electrons collide on the lead conversion target. Even though the PMT was in a gate-off state, it still generates a substantial electrical pulse output, marking the time of muon generation, $t_{start}$. The yellow curve in the figure represents the typical raw data recorded by the oscilloscope. When the MCP-PMT is gated on, there is a discernible jump in the current output curve, followed by numerous sharp peaks along the curve. Figure 2(b) displays the waveforms of several typical peaks, each signifying a muon decay event. The oscilloscope-recorded waveforms exhibit a complex baseline, significant background noise, and random pulse amplitudes, posing challenges for the precise identification of muon decay events. To address this, we employ a SNIP (Sliding Normalization and Interval Probability) baseline subtraction method in conjunction with a continuous wavelet transform peak-finding algorithm (see Method). Initially, we analyze the baseline of Figure 2(a), resulting in the black curve displayed in the figure. Subtracting this baseline (black curve) from the raw data (yellow curve) yields the waveform identified as a muon decay event (blue curve). Each green vertical line in Figure 2(a) indicates a muon decay event, $t_{stop}$, enabling the determination of an individual muon's lifetime. By accumulating data from multiple shots, a distribution of muon decay lifetimes can be established. The varying amplitudes of each peak arise from the different energies deposited in the liquid scintillators. Figure 2(c) and 2(d) show the two-dimensional scatter plot of the signal amplitude out of the MCP-PMT versus time for Detector 1# and Detector 2#. A noticeable decay trend in the number of muon events is observed. Typically, the distribution of amplitudes is unrelated to the muon decay time; however, the scatter plot reveals that in the early stages of decay, the amplitudes are higher compared to the later stages. This disparity arises from the higher rate of muon decay events initially, potentially leading to multiple decays occurring within a short timeframe. If two decay events occur within the scintillator's afterglow decay time, they might be erroneously counted as a single event, with the output current amplitude representing the sum of both decays, resulting in a higher average amplitude for early decay events. Figures 2(e) and 2(f) demonstrate the average signal amplitude per unit time as a function of the $\Delta t$. It is evident that the average amplitude of muon events in the early decay stages is 2 to 3 times higher than in the later stages. However, the average signal amplitude of muon events converges over time after 5 μs. Therefore the distribution of muon with $\Delta t > 5 \mu s$ was fitted to obtain the lifetime of muons.

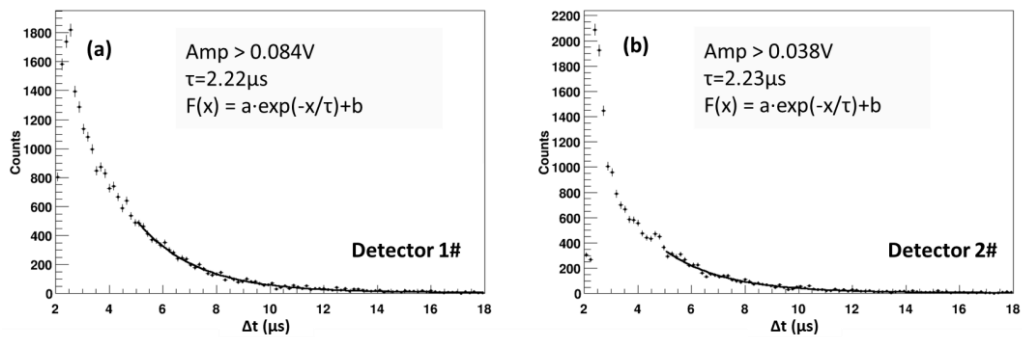

**Figure 3 Distributions of the number of muon decay events over time for Detector 1# (a) and Detector 2# (b) exhibits excellent exponential distributions consistent with the muon's lifetime.**

Figure 3 displays the distribution of muon decay events over $\Delta t$ in Detector 1# and Detector 2#, showcasing a remarkable exponential distribution. Further exponential fitting for the time interval between 5 μs and 18 μs yields a result of 2.2 μs, consistent with the muon's lifetime, validating the physical signals generated by muons. The figure also indicates that with an accumulation of 178 shots, a total of $2.6\times10^4$ muon events were recorded from Detector 1# ($2.0\times10^4$ from Detector 2#). Furthermore, it is necessary to perform a global Geant4 simulations based on these findings to model muon production and detector response accurately, enabling the derivation of detailed information such as the actual flux of produced muons.

The Geant4 package is utilized to replicate the generation and detection processes of muons. For simulation simplicity, a mono-energetic electron beam is set at 0.6 GeV, 0.8 GeV, 1.0 GeV, 1.2 GeV, and 1.4 GeV respectively. The detailed setup of the Geant4 simulation was elaborated on in Method. The simulation accounts for two primary mechanisms: muon production via pion decays (including photo-production and electro-production) and muon pair production through the Bethe-Heitler process. Subsequently, the simulated pions/muons are injected into a Geant4 detector model, encompassing the lead conversion target, lead shielding, and the liquid scintillator detectors employed in the experiment.

In the simulation, we initially analyzed the energy spectrum and angular distribution of pions exiting the conversion target, as shown in Figure 4 (a) and (c). It is evident that as the incident electron energy increases, on one hand, the energy distribution and peak energy of the pions almost keep constant, and on the other hand, the number of produced π quickly rises. The distribution of pions leaving the conversion target displays a distinct near-4π distribution, with pions radiating in forward, lateral, and backward directions relative to the electron beam. This phenomenon is attributed to the characteristics of the photo-production and electro-production processes, as well as the scattering of pions within the conversion target. The extensive lead shielding walls utilized in the experiment effectively enhance the detection efficiency of these isotropically distributed muons.

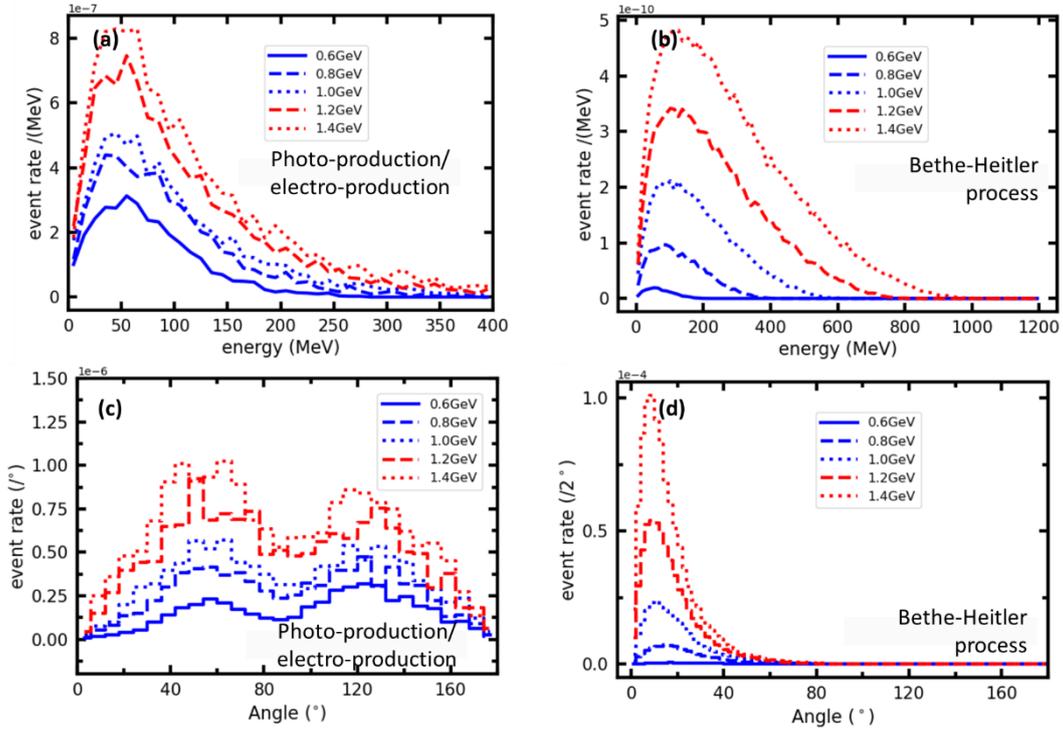

**Figure 4 The energy spectrum (a) and angular distribution (c) of $\pi$ out of the Pb converter in photo-production and electro-production; (b) and (d) of muon pair out of the Pb converter in Bethe-Heitler process normalized to a single incident electron.**

Furthermore, we explored the energy spectra and angular distributions of muon pairs generated through the Bethe-Heitler process as shown in Figure 4 (b) and (d). It is apparent that with an increase in the energy of the incident electrons, both the peak energy and the overall distribution of the muons exhibit significant growth which is attributed to the heightened cross sections of the Bethe-Heitler process corresponding to the energy of the incident electrons. Additionally, the angular distribution of the outgoing muon pairs displays a clear inclination to align with the direction of the incident electrons. As the energy of the incident electrons rises, the emission of muon pairs becomes more collimated, facilitating the generation of muon beams with reduced emittance and higher energy levels.

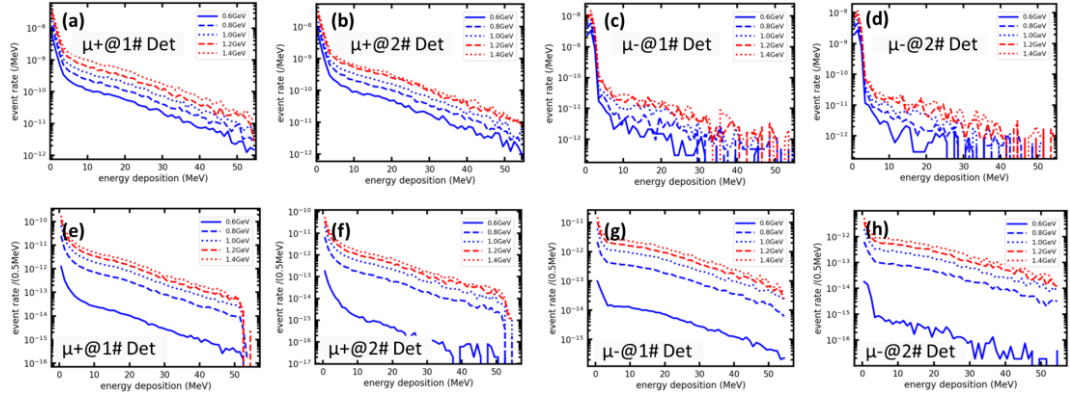

**Figure 5 Energy deposition of muons from photo-production/electro-production (a)~(d) and Bethe-Heitler process (e)~(h) normalized to a single incident electron.**

Subsequently, the pions and muon pairs were transported into the simulated detectors to evaluate the resulted energy deposition, as illustrated in Figure 5. The figures highlight that among the events detected by the current detectors, those originating from the photo-production/electro-production process predominate, while events from the Bethe-Heitler process are relatively rare. This discrepancy arises from the notably larger cross section of the former, which exceeds that of the latter by three to four orders of magnitude. Furthermore, given that the angular distribution of pions generated by the photo-production process closely aligns with a near-4π solid angle distribution, the number of events detected by Detector 1# and Detector 2# is approximately equal. Moreover, only about 2%~3% of the detected events are come from $\mu^-$ event stopped inside the detectors. Most of $\mu^-$ event are captured due to the capture effect within the lead shielding walls. However, it is essential to note that these captured $\mu^-$ undergo a modified (shortened) decay lifetime but do not impact the measurement of the muon lifetime as the corresponding $\Delta t$ fall outside the detection range.

By using the Geant4 simulation package, we conducted an in-depth exploration of all potential final states resulting from high-energy electron beams colliding with the lead converter target. The primary objective was to ascertain the presence of any inadvertent coincidence events stemming from processes like nuclear excitation. The simulation encompassed a vast scale of events, totaling $9\times10^9$, and upon meticulous consideration of all possible processes. Our simulation studies indicate that some low-energy photonuclear neutrons may be detected by liquid scintillator detectors on the microsecond timescale. However, the energy deposited by these neutrons is significantly lower than the experimental threshold of the detector. Therefore, they do not affect the measurement of the muon event. Additionally, a limited number of events exhibit higher energy deposits in the detector. Subsequent tracking of these events revealed that their genesis consistently involved muons, thereby categorizing them as genuine muon events. This simulation effectively ruled out the influence of other processes such as nuclear excitation, ensuring the focus on muon-related phenomena.

This simulation also fully demonstrates that employing a delayed trigger combined with lifetime measurement was a robust methodology for discerning transient high-flux muons engendered by ultra-intense ultrafast lasers. Furthermore, we deliberated on the potential impact of accidental coincidence events originating from cosmic rays. By factoring in the flux of cosmic ray muon events in conjunction with the detector volume, the estimated number of effective cosmic ray events during a single detection period is less than 10⁻³, rendering their influence negligible and warranting no significant consideration.

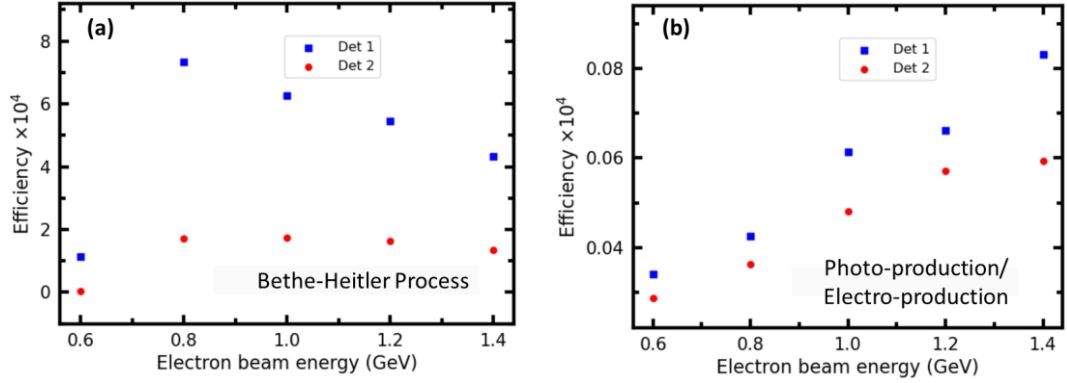

**Figure 6 Detection efficiencies of muons from (a) Bethe-Heitler Process and (b) Photo-Production**

The results of the detection efficiency analysis utilizing the current muon detector setup, derived from Geant4 simulations, are visually represented in Figure 6. Notably, the detection efficiency for muons generated through the Bethe-Heitler process surpasses that of the photo-production/electro-production by approximately two orders of magnitude for the higher energy of muons. Furthermore, concerning muons originating from the Bethe-Heitler process, the efficiency of Detector 1 markedly exceeds that of Detector 2, primarily due to the former's proximity to the direction of the electron beam. It is also worth highlighting that despite variations in detection efficiency for different incident electron energies, the fluctuations are within a few-fold range, enabling the averaging of detection efficiencies across various incident energy points to estimate the overall detection efficiency for a diverse spectrum of electron beams in the experimental setup. Consequently, the deduced muon detection efficiency for this experiment stands at approximately $5\times10^{-6}$. It is also shown that only 25% pion from the photo-production/electro-production processes can get out of the converter target. Considering the yield of the detected muon in the experiment, it becomes apparent that the muon yield out of the converter target in this setup reaches the order of $10^9$, translating to a single-shot muon yield ranging to $10^7$, indicative of a significantly high-yield muon source.

In summary, for the first time we have successfully established a clear demonstration of laser produced muon within a laser laboratory, achieving a notable muon yield of 0.01μ per electron at a central energy around 1 GeV. This groundbreaking muon source boasts high instantaneous flux and short pulses, marking a significant advancement in muon production capabilities. Presently, commercially available 200 TW-level repetitive laser systems demonstrate the ability to efficiently accelerate electrons to 1 GeV utilizing laser wakefield acceleration, yielding single-shot electron charges on the order of $10^8$ and offering repetition rates of up to 10 Hz. Leveraging these cutting-edge laser technologies, it becomes feasible to generate pulsed muon sources with yields of $10^7$/s and pulse widths of tens of nanosecond within compact laser laboratories, thereby lowering the technical barriers for muon application technologies such as μSR or muonic X-rays and enhancing the efficacy of muon-related applications. This innovation also presents opportunities for utilizing pulsed, high-flux muon sources as injection sources for muon colliders, further expanding the scope of muon research and applications.

Moreover, the electron energies achievable through laser wakefield acceleration can extend to the 10 GeV level, resulting in the production of muon pairs with energies reaching several GeV, thereby paving the way for the realization of muon imaging within shorter timeframes [38]. With the rapid development of ultra-short pulse laser technology, laser-driven muon sources are expected to become a novel type of radiation source, playing a significant role in various research fields.

accelerator, Plasma Phys. Control. Fusion, 60, 095002, (2018)

34. K. Nagamine, H. Miyadera, A. Jason, R. Seki, Compact muon source with electron accelerator for a mobile μSR facility, Physica B, 404, 1020–1023 (2009)

35. Feng Zhang, Boyuan Li, Lianqiang Shan, Bo Zhang, Wei Hong, and Yuqiu Gu, A new method on diagnostics of muons produced by a short pulse laser, High Power Laser Sci. Eng. , Vol. 5, e16, (2017)

36. J. Allison et al., "Geant4 developments and applications," in IEEE Transactions on Nuclear Science, 53(1) , 270-278, (2006). doi: 10.1109/TNS.2006.869826.

37. Zongxin Zhang, Fenxiang Wu, Jiabing Hu, et al., The 1 PW/0.1 Hz laser beamline in SULF facility, High Power Laser Sci. Eng. 8, e4 (2020).

38. Luke Calvin, Paolo Tomassini, Domenico Doria, Daniele Martello, Robert M. Deas and Gianluca Sarri, Laser-driven muon production for material inspection and imaging, Front. Phys. 11:1177486. (2023) doi: 10.3389/fphy.2023.1177486.**Methods**

**Electron Acceleration Experiment Setup.**

The experiment was conducted at the Shanghai Super-intense Ultrafast Laser Facility (SULF) [37], where a petawatt laser system was utilized. This advanced system incorporates dual-chirped pulse amplification technology along with a cascaded nonlinear pulse cleaning approach. Capable of generating high-contrast femtosecond laser pulses with peak powers scaling to the petawatt level and a central wavelength of approximately 800 nm, the system operates at a repetition rate of 0.1 Hz. These femtosecond laser pulses, with a duration of around 30 femtoseconds and the ability to attain a focused intensity of up to $10^{20}$ W/cm$^2$, function as the primary light source for a diverse range of intense field physical science investigations, including electron acceleration, proton acceleration, and the development of secondary radiation sources.

During the experiment, the femtosecond laser pulses were focused by an off-axis parabolic mirror with an $f$/60 F-number, enabling the peak power density to reach (1.5-2)×$10^{19}$ W/cm$^2$, corresponding to a laser normalized intensity denoted by $a_0 \approx$ 2.5-2.9. Subsequently, the laser was directed into a high-density pure helium gas jet produced by an elliptical nozzle with a long axis measuring 8.8 mm and a short axis of 4.5 mm. The gas pressure was meticulously adjusted to range between 7-10 bar, resulting in a plasma density of (0.7-1)×$10^{19}$ cm$^{-3}$, thereby facilitating the generation of GeV-level high-energy electron beams.

**Detector setup and muon detection energy threshold calibration.**

In the experimental setup, a liquid scintillator detector was employed to capture electron/positron signals from muon decay. This cylindrical detector boasts an outer diameter of 15.2 cm and a thickness of 11.2 cm, enclosed within an aluminum shell measuring 0.6 cm in thickness. The interior of the detector is filled with liquid scintillator predominantly comprised of PX (p-xylene) serving as the solvent, alongside PPO and bis-MSB scintillators as solutes. When electron/positron signals from muons interact within the scintillator, visible light is generated and transmitted through optical glass before being detected by an MCP-PMT, which converts the optical signals into electrical pulses subsequently recorded by an oscilloscope.

To establish a clear correspondence between the amplitude of the electric pulse signals emitted by the MCP-PMT and the deposited energy, the detector's sensitivity can be effectively calibrated utilizing cosmic rays. Firstly, by adopting the same detector settings as in the experiment, we obtained the spectrum of current signal strengths output by cosmic rays in the liquid detector. Utilizing Geant4 simulations, it is also possible to study the energy deposition process of cosmic rays in the detectors, thereby deriving the energy deposition spectrum. The relationship between the detector's current signal strength and energy deposition inside the detectors can be established straightforwardly. Ultimately, we determined that the corresponding energy detection threshold for the detector used in the experiment was around 18 MeV.

**Data processing**

The measured signal contains various components, including the electromagnetic interference signal, the electronic noise, the pulse signal of muons and the baseline contributed by the DC component of the readout circuit and pulse accumulation. To identify the randomly distributed muon decay pulses and extract the pulse starting time from the measured signal, a preprocessing algorithm was designed. This algorithm comprises convolution filtering, wideband background correction, and continuous wavelet transform.

The convolution filtering was implemented using a 13-point convolution window to handle high-frequency electronic noise. Wideband background correction was accomplished using a sensitive nonlinear peak clipping (SNIP) algorithm. The SNIP algorithm is an iterative background elimination method widely employed in γ-ray spectrum processing [39]. In this work, the iteration time was set to 80 to address the baseline.

Given that the pulse amplitudes and widths of the muon signal may vary, a continuous wavelet transform (CWT)-based peak detection algorithm was utilized. This algorithm can identify peaks with different scales and amplitudes [40]. The CWT-based peak detection algorithm detects peaks by employing a shape-matching function that provides a "goodness of fit" coefficient. In this work, the shape-matching function used was the Ricker wavelet. The widths used for calculating the CWT matrix ranged from 1 to 20 with increments of 0.5. This range was chosen to cover the expected widths of the peaks of interest. The peak positions identified by the CWT-based peak detection algorithm were considered as the starting times of the muon signal pulses.

**Geant4 simulation setup**

We utilized the Geant4 simulation package to model the production of muons from monoenergetic electron beams and their detection process. The conversion target, lead shielding, and liquid scintillator detector used in the experiment were all modeled. Since the cross section for direct muon production by electrons is relatively low, to avoid computing a large number of electromagnetic shower processes unrelated to muons, the simulation was divided into two steps: The first step involved simulating the production of pions through photo-production/electro-production in the conversion target and the production of muon pairs through the Bethe-Heitler process. For this purpose, a hemispherical vacuum ideal detector was placed around the conversion target to record the momentum and energy information of the produced pions and muon pairs. In the next step, these pions and muons were then inserted into a fully physically modeled detector to simulate detection efficiency.

In the simulation, we used the QGSP_BERT physics list and activated the *GammaToMuPair* reaction. Considering that the cross-section for photo-production/electro-production is about 3 to 4 orders of magnitude higher than that of the Bethe-Heitler process, the Bethe-Heitler process cross-section was artificially increased by a factor of 10000 in the simulation by modifying the *gmumuFactor* value to 10000.

In the next step, using the information of the produced pion and muon pairs, we can estimate the energy deposition spectrum and the number of effective signals in the detector through simulation. We sample the energy and momentum direction of the pions/muons collected by the ideal detector, set these as initial particles, and place them into the well-modeled detector to obtain the energy deposition spectrum and time information for charged particles. It is noteworthy that even if multiple tracks enter the detector and deposit energy within the same event, their time intervals are extremely short (not exceeding nanoseconds). Therefore, the actual energy deposition spectrum of a single event is statistically summarized based on the total for each track. By applying the same energy threshold and time range as in the experiment the detection efficiencies can be obtained as shown in Figure 6.

**Data availability**

Data supporting the findings of this study are available from the corresponding authors upon reasonable request.

**Code availability**

Geant4 code and the SNIP algorithm supporting the findings of this study are available from the corresponding authors upon reasonable request.

**Acknowledgments**

We thank all the Shanghai Super-intense Ultrafast Laser Facility (SULF) technical staff for their support during the experiment. The research leading to these results has received funding from the Science Challenge Project TZ2024016, the National Natural Science Foundation of China (grants nos.12175209, 12235014, 11805182, 12105327, 12105353, 12388102)，National Key R&D Program of China Grant No.

2022YFA1603300, the Guangdong Basic and Applied Basic Research Foundation (Grant No. 2023B1515120067) and the CAS Project for Young Scientists in Basic Research (YSBR060).

**Author contributions**

F.Z, W.M.Z., Y.Q.G., W.T.W., Y.X.L., Z.Y.S. and R.X.L. conceived this project, which was designed by F.Z., W.T.W., and L.W.C.. The experiment was carried out by F.Z., Y.J.G., B.C., K.F., H.W., H.J.L., Z.G.D., L.Q.S., Z.Q.Y., C.T., Z.X.Z., J.Y.Q., J.C.Z. and Y.X.. The data was analyzed by F.Z. J.X.W. and B.C., the numerical simulations were performed by F.Z., L.D., L.W.C., C.W., Z.W.P., D.Y., Y.H.Y., X.H.Z., L.Y.. The paper was written by F.Z., L.D., L.W.C. and W.T.W.

**Competing interests**

The authors declare no competing interests.

**Correspondence and requests for materials** should be addressed to L.W.C., W.M.Z., Y.Q.G and W.T.W.